\documentclass[a4paper,english,10pt,amssymb,nofootinbib,superscriptaddress]{jpconf}
\usepackage{lmodern}
\usepackage{lmodern}
\usepackage[T1]{fontenc}
\usepackage[utf8]{inputenc}
\usepackage{babel}
\usepackage{verbatim}
\usepackage{bm}
\usepackage{amsmath}
\usepackage{amsthm}
\usepackage{amssymb}
\usepackage{physics}
\usepackage{esint}
\usepackage{cite}
\usepackage[unicode=true,pdfusetitle,
 bookmarks=true,bookmarksnumbered=false,bookmarksopen=false,
 breaklinks=false,pdfborder={0 0 1},backref=false,colorlinks=false]
 {hyperref}

\makeatletter

\pdfpageheight\paperheight
\pdfpagewidth\paperwidth

\usepackage{babel}
\usepackage{amsthm}\usepackage{bm}\usepackage{verbatim}\usepackage[most]{tcolorbox}

\theoremstyle{plain} 

\usepackage{bm}

\makeatother

\numberwithin{equation}{section}

\renewcommand{\dd}{\mathrm{d}}

\newcommand{\eps}{\varepsilon}

\newcommand{\AdS}{\mathrm{AdS}}
\newcommand{\dS}{\mathrm{dS}}
\newcommand{\ISO}{\mathrm{ISO}}
\newcommand{\SO}{\mathrm{SO}}

\newtheorem{prop}{Proposition}[section]
\newtheorem{remark}{Remark}[section]

\begin{document}
\title{\bf The Flat Critical Branch Between Nariai and Bertotti--Robinson Geometries as a Solution of Cosmological Einstein--Maxwell Theory \\}

\author{Metin G{ü}rses}
\address{Department of Mathematics, Bilkent University, 06800 Ankara, T{ü}rkiye}
\ead{gurses@fen.bilkent.edu.tr}

\author{Tahsin Çağr{ı} Şişman}
\address{Department of Astronautical Engineering, University of Turkish Aeronautical Association, 06790 Ankara, T{ü}rkiye}
\ead{tahsin.c.sisman@gmail.com}

\author{Bayram Tekin}
\address{Department of Physics, Bilkent University, 06800 Ankara, T{ü}rkiye}
\ead{bayram.tekin@bilkent.edu.tr}

\begin{center}
{\small \itshape
To appear in the proceedings of \emph{60 Years in Mathematical Physics: Gürses-Fest} issue of
\emph{Journal of Physics: Conference Series} (JPCS, IOP Publishing).}
\end{center}

\begin{abstract}
We analyze a class of product geometries of the form $\mathbb{R}^{1,1}\times \Sigma_2$ supported by electric, magnetic, or dyonic flux in the Einstein--Maxwell–$\Lambda$  theory. These spacetimes belong to a unified family of direct products $(\dS_2,\mathbb{R}^{1,1},\AdS_2)\times \Sigma_2$ distinguished solely by the sign of the Lorentzian curvature of the two–dimensional factor. We focus on the critical configuration for which the Lorentzian curvature vanishes. At this balance point between the cosmological curvature and the Maxwell stress, the longitudinal geometry becomes exactly flat while the transverse sphere radius is fixed algebraically by the conserved flux. We refer to this geometry as the \emph{critical Maxwell flux string}: a homogeneous flux-supported geometry curved only in the transverse directions and invariant along a two–dimensional null worldvolume. It represents the algebraic midpoint interpolating between the Nariai $(\dS_2\times S^2)$ and Bertotti–Robinson $(\AdS_2\times S^2)$ spacetimes. A qualitative structural change occurs precisely at this midpoint. The spacetime is Petrov type~D with constant scalar curvature invariants, placing it in the degenerate Kundt/CSI class. Because the curvature structure reduces any polynomial rank-two tensor to a linear combination of the metric and the Maxwell stress tensor, the same configuration solves a broad class of algebraic higher-curvature gravity theories. In this sense, the critical flux string and its aligned deformations constitute \emph{almost universal} solutions.    
\end{abstract}

\section{Introduction}

Flux-supported product spacetimes occupy a central position in classical and quantum gravity. Among the simplest exact solutions of the Einstein–Maxwell system are direct products of constant-curvature two-manifolds, most notably the Bertotti–Robinson geometry $\AdS_2\times S^2$ \cite{Bertotti1959,Robinson1959} and the Nariai universe $\dS_2\times S^2$ \cite{Nariai1950,Nariai1951}. The former arises as the near-horizon limit of extremal Reissner--Nordstr\"om black holes and underlies the attractor mechanism \cite{ferrara-kallosh-strominger,sen-attractor}, while the latter appears as the maximal black-hole limit in de~Sitter space \cite{GinspargPerry1983}. More generally, such product geometries play an important role in near-horizon classifications \cite{kunduri-lucietti} and in flux compactifications following the Freund–Rubin mechanism \cite{freund-rubin}. Furthermore, product geometries in the Lovelock gravity setting were studied in \cite{Kastor-Senturk} as the symmetry breaking vacua.

A key feature of these solutions is their algebraic simplicity: the Riemann tensor is completely determined by a small number of curvature scales, and the Einstein–Maxwell equations reduce to algebraic relations among those scales and the conserved flux. In four dimensions, this naturally leads to a unified family of product spacetimes of the form
\[
(\dS_2,\mathbb{R}^{1,1},\AdS_2)\times S^2,
\]
distinguished solely by the sign of the Lorentzian (Gaussian) curvature $K_L$ of the two-dimensional Lorentzian factor which we called as the longitudinal sector while we called the remaining $S^2$ space as the transverse sector.

In this work, we identify and analyze the \emph{critical} member of this family: a homogeneous $\mathbb{R}^{1,1}\times S^2$ background supported by a nonzero Maxwell flux and a nonzero cosmological constant. Its defining property is that the Lorentzian factor is exactly flat. At this point, the cosmological curvature and the Maxwell stress cancel in the longitudinal sector, while the transverse sphere radius is fixed algebraically by the conserved charge. In terms of the unified family, this branch corresponds precisely to the vanishing of the Gaussian curvature
\[
K_L=0,
\]
the midpoint between the Nariai ($K_L>0$) and Bertotti–Robinson ($K_L<0$) regimes.

The physical interpretation changes qualitatively at this midpoint. The $\dS_2\times S^2$ and $\AdS_2\times S^2$ branches possess enhanced $\SO(1,2)$ symmetry in the Lorentzian sector and are closely tied to extremal black-hole near-horizon limits. By contrast, the critical branch has $\ISO(1,1)\times\SO(3)$ symmetry and is best interpreted not as a black-hole geometry but as a homogeneous flux geometry--a four-dimensional \emph{flux string} or fluxbrane. Similar flux-supported configurations and their higher-dimensional generalizations have been studied extensively in gravitational theories \cite{freund-rubin,melvin,gurses92,gur-ser}.

A further distinction appears when null deformations of the background metric are considered. In Brinkmann form \cite{Brinkmann1925}, the $uu$ Einstein equation for the critical branch degenerates into a transverse Laplace equation. This permits formally free wave profiles propagating along the null direction and places the geometry within the Kundt class \cite{Kundt1961,griffiths-podolsky}. Such a degeneracy is absent in the curved $\dS_2\times S^2$ and $\AdS_2\times S^2$ branches, where the longitudinal curvature fixes the geometry algebraically. For a round $S^2$, global regularity eliminates nontrivial smooth harmonic profiles, restoring rigidity; nevertheless, the existence of the degeneracy itself is a structural signature of the flat midpoint. Related Kundt waves on direct-product electro-vacuum backgrounds, including the Pleba\'nski--Hacyan universes, have been analyzed in \cite{PlebanskiHacyan1979,Kadlecova2009}.

The curvature structure of the critical background is also highly constrained. The undeformed geometry is Petrov type~D with principal null directions aligned along the longitudinal factor, and all scalar curvature invariants are constant, placing the spacetime within the CSI framework \cite{coley-csi}. Because the traceless Ricci tensor is aligned and the Riemann tensor contains only a small number of independent scalars, any symmetric rank-two tensor constructed polynomially from the curvature reduces to a linear combination of the metric and the Maxwell stress tensor. Consequently, the same configuration solves a broad class of higher-curvature theories, including $f(\mathrm{Riemann})$ and quadratic gravity, once the same algebraic tuning conditions are imposed. In this precise sense, the critical flux string belongs to the class of \emph{almost universal spacetimes} \cite{HervikPravdaPravdova2017,kuchynka-universal}.

The present work forms part of a series on almost universal Kerr--Schild families of metrics. In \cite{GST-AUM2026} we showed that the null deformation, which we called constant-curvature $pp$-waves, with $\mathbb{R}^{1,1}\times S^{2}$ or $\mathbb{R}^{1,1}\times H^{2}$ backgrounds are almost universal and solve the cosmological Einstein--Maxwell equations, realizing the flat midpoint between the Nariai and Bertotti--Robinson branches for the $\mathbb{R}^{1,1}\times S^{2}$ topology. In the companion work \cite{GST-KSK2026} we generalize the Kerr--Schild-Kundt (KSK) construction to these constant-curvature $pp$-waves, obtaining almost universal solutions across broad classes of higher-curvature theories. Before this work, the KSK family of metrics was formed by the (A)dS-waves \cite{gst4}. The (A)dS-waves are almost universal metrics \cite{gst4, ghst, gst2}, which were first found to be the solutions of quadratic gravity \cite{ggst, gst1}. Therefore, the constant-curvature $pp$-waves and the (A)dS-waves are closely related. The (A)dS-waves have higher-dimensional \cite{gst4, ghst, gst2} and lower-dimensional counterparts \cite{gst3}. Thus, it is natural to expect similar higher- and lower dimensional generalizations for both the background $\mathbb{R}^{1,1}\times \Sigma_2$ product metric and its constant-curvature $pp$-wave deformation. Furthermore, the (A)dS-waves can be obtained from the smooth curves of codimension 1 space or spacetime \cite{gst5}. As a future outlook, a similar construction for the constant-curvature $pp$-waves may enlarge the KSK class with new almost universal metrics.

Although the geometry itself is simple, its role within the $(\dS_2,\mathbb{R}^{1,1},\AdS_2)\times S^2$ family is subtle.
By treating the problem from first principles, we clarify the special status of the $K_L=0$ branch as a geometric midpoint combining algebraic rigidity and almost universality across a wide class of metric-based gravitational theories.

The paper is organized as follows. We first derive the geometric properties of the $\mathbb{R}^{1,1}\times\Sigma_2$ product metric and compute its curvature tensors explicitly. We then determine the Petrov type and Newman–Penrose scalars, solve the Einstein–Maxwell equations algebraically, and finally analyze aligned Brinkmann-type deformations. Our goal is not merely to exhibit another product solution but to clarify the structural role of the $K_L=0$ branch as the algebraic transition point between the Nariai and Bertotti–Robinson geometries.

\section{The product metric and basic geometry}

We now introduce the geometric ansatz underlying the critical branch. Throughout this work, we consider a four-dimensional spacetime of direct-product form
\begin{equation}
\mathcal{M}_4 = \mathbb{R}^{1,1} \times \Sigma_2 ,
\label{M4product}
\end{equation}
equipped with the metric
\begin{equation}
ds^2 = 2\,du\,dv + h_{ab}(x^c)\,dx^a dx^b,
\qquad a,b,c=2,3.
\label{productmetric}
\end{equation}
Here $(u,v)$ are double-null coordinates on the two-dimensional Lorentzian factor, and $h_{ab}(x^c)$ is an arbitrary Riemannian metric on the transverse two-manifold $\Sigma_2$. The metric is independent of $u$ and $v$.

This ansatz captures the $K_L=0$ branch of the $(\dS_2,\mathbb{R}^{1,1},\AdS_2)\times S^2$ family. Unlike the curved branches, the longitudinal sector is exactly flat, and all nontrivial curvature resides entirely in the transverse geometry.

\subsection{Coordinate conventions and inverse metric}

The only nonvanishing components of the metric are
\begin{equation}
g_{uv}=g_{vu}=1,
\qquad
g_{ab}=h_{ab}(x^c),
\end{equation}
with inverse
\begin{equation}
g^{uv}=g^{vu}=1,
\qquad
g^{ab}=h^{ab}(x^c),
\end{equation}
and
\[
g_{uu}=g_{vv}=g^{uu}=g^{vv}=0.
\]
With these metric components, the two immediate geometric consequences are as follows:
\begin{itemize}
\item[(i)] The vector fields $\partial_u$ and $\partial_v$ are null and mutually orthogonal.
\item[(ii)] Because the metric components are independent of $u$ and $v$, and no mixed longitudinal--transverse terms appear, these vectors will turn out to be covariantly constant.
\end{itemize}

Thus the longitudinal sector admits a pair of aligned null directions, a property that will play a central role in the Petrov classification and the Kundt structure discussed later.

The block structure of the metric implies a complete decoupling between longitudinal and transverse geometry. A direct computation of the Christoffel symbols shows that
\begin{itemize}
\item all Christoffel symbols with at least one index in $\{u,v\}$ vanish;
\item the only nonzero connection coefficients are the intrinsic Christoffel symbols of $(\Sigma_2,h)$:
\begin{equation}
\Gamma^c_{ab}
=
\frac12
h^{cd}
\left(
\partial_a h_{bd}
+\partial_b h_{ad}
-\partial_d h_{ab}
\right).
\label{GammaSigma}
\end{equation}
\end{itemize}
In particular, the longitudinal null vectors are covariantly constant:
\begin{equation}
\nabla_\mu (\partial_u)^\nu = 0,
\qquad
\nabla_\mu (\partial_v)^\nu = 0.
\label{covconstant}
\end{equation}
The existence of two covariantly constant null vectors is a highly restrictive property. It places the spacetime within the class of degenerate Kundt geometries \cite{griffiths-podolsky}, and implies that the longitudinal direction defines a preferred null congruence.
\section{Curvature of the product spacetime}

We now compute the curvature of the product spacetime $\mathcal{M}_4 = \mathbb{R}^{1,1} \times \Sigma_2$ equipped with the metric \eqref{productmetric}. Throughout this section $R^{(2)}_{abcd}$, $R^{(2)}_{ab}$, and $R^{(2)}$ denote curvature tensors constructed solely from the transverse geometry $(\Sigma_2,h)$.

\subsection{Two-dimensional curvature identities}

A key simplification arises from the fact that, in two dimensions, the Riemann tensor possesses only a single independent component. It is completely determined by the scalar curvature:
\begin{equation}
R^{(2)}_{abcd}
=\frac{R^{(2)}}{2}\,
\big(h_{ac}h_{bd}-h_{ad}h_{bc}\big).
\label{eq:2d_Riemann_identity}
\end{equation}
Contracting \eqref{eq:2d_Riemann_identity} immediately yields
\begin{equation}
R^{(2)}_{ab}
=\frac{R^{(2)}}{2}\,h_{ab}.
\label{eq:2d_Ricci_identity}
\end{equation}
Thus, once the scalar curvature of $\Sigma_2$ is specified, its entire curvature tensor is fixed. This rigidity of two-dimensional geometry will propagate directly into the four-dimensional curvature structure.

\subsection{Four-dimensional Riemann, Ricci, and scalar curvature}

Because the Levi--Civita connection splits completely between longitudinal and transverse sectors, any Riemann component with at least one $u$ or $v$ index vanishes. The only nonzero components are purely transverse:
\begin{equation}
R_{abcd} = R^{(2)}_{abcd}.
\label{eq:4d_Riemann_equals_2d}
\end{equation}
The Ricci tensor therefore has support only in the transverse block:
\begin{equation}
R_{ab}=\frac{R^{(2)}}{2}\,h_{ab},
\qquad
R_{uu}=R_{vv}=R_{uv}=R_{ua}=R_{va}=0.
\label{eq:4d_Ricci}
\end{equation}
The scalar curvature reduces identically to that of $\Sigma_2$:
\begin{equation}
R=g^{\mu\nu}R_{\mu\nu}
= h^{ab}R_{ab}
= R^{(2)}.
\label{eq:4d_scalar_equals_2d}
\end{equation}

\medskip

\noindent
\emph{Geometric interpretation.}
All curvature resides entirely in the transverse two-manifold $(\Sigma_2,h)$, while the longitudinal $\mathbb{R}^{1,1}$ factor
remains intrinsically flat. The Riemann tensor, therefore, splits into a purely transverse curvature block, with all mixed components vanishing. This block structure underlies the algebraic simplicity of the Einstein equations for the product geometry.

\subsection{Einstein tensor}

Using
\begin{equation}
G_{\mu\nu}
=R_{\mu\nu}
-\frac{1}{2}Rg_{\mu\nu},
\label{eq:Einstein_def}
\end{equation}
we obtain
\begin{align}
G_{ab}
&=\frac{R^{(2)}}{2}h_{ab}
-\frac{1}{2}R^{(2)}h_{ab}
=0,
\label{eq:Gab_zero}\\[4pt]
G_{uv}
&=R_{uv}
-\frac{1}{2}R g_{uv}
= -\frac{1}{2}R^{(2)},
\label{eq:Guv}\\[4pt]
G_{uu}
&=G_{vv}=G_{ua}=G_{va}=0.
\label{eq:G_other_zero}
\end{align}
Equation \eqref{eq:Gab_zero} is a purely \emph{kinematical} consequence of the product ansatz: the transverse Einstein tensor vanishes identically, independently of the choice of $h_{ab}(x)$.

This fact has an important structural implication. Since $G_{ab}=0$ identically, the transverse Einstein equations reduce entirely to algebraic conditions once matter sources are specified. In particular, when a homogeneous Maxwell flux is introduced, the balance between transverse curvature, flux energy, and cosmological constant determines the sphere radius algebraically.
\section{Weyl tensor, Newman--Penrose scalars, and Petrov type}

We now determine the algebraic type of the product geometry. Because the longitudinal sector admits two covariantly constant null vectors, the Newman--Penrose formalism is particularly well adapted to this background.

\subsection{Null tetrad adapted to the product structure}

We introduce a Newman--Penrose null tetrad adapted to the product structure $\mathcal{M}_4=\mathbb{R}^{1,1}\times \Sigma_2$. Let the spacetime coordinates be denoted by $x^\mu = (u, v, x^2, x^3)$. The two real null legs spanning the longitudinal factor are taken along the null coordinate directions as
\begin{align}
\ell^\mu &= \delta^\mu_{v}, 
\qquad
n^\mu = \delta^\mu_{u},
\qquad
\ell\cdot n = 1 .
\label{eq:NP_ln}
\end{align}
The two complex null legs are constructed from the orthonormal zweibein $\{e^a_{(1)}, e^a_{(2)}\}$ of $(\Sigma_2,h)$, with $a\in\{2,3\}$, satisfying
\[
h_{ab}\,e^a_{(A)}\,e^b_{(B)} = \delta_{AB}, \qquad A,B\in\{1,2\}.
\]
Embedding into the full spacetime by setting the longitudinal components to zero gives
\begin{equation}
e_{(A)}^\mu = (0,0,e_{(A)}^{\,2},e_{(A)}^{\,3}) .
\end{equation}
The complex null vectors in the transverse directions are then defined by
\begin{align}
m^\mu &= \frac{1}{\sqrt{2}}\left(e_{(1)}^\mu + i\,e_{(2)}^\mu\right),
\qquad
\bar m^\mu = \frac{1}{\sqrt{2}}\left(e_{(1)}^\mu - i\,e_{(2)}^\mu\right),
\label{eq:NP_m}
\end{align}
where nullity $m\cdot m =0$ follows from the orthonormality condition $\delta_{AB}$. These vectors satisfy the standard Newman--Penrose inner-product relations
\begin{equation}
m\cdot \bar m = 1,
\qquad
\ell\cdot m = 0,
\qquad
n\cdot m = 0 .
\end{equation}
Thus, the tetrad is naturally adapted to the product geometry: $\ell$ and $n$ span the longitudinal $\mathbb{R}^{1,1}$ factor, while $m$ and $\bar {m}$ lie entirely within the tangent space of $\Sigma_2$.

\subsection{Weyl decomposition}

We now compute the Weyl tensor of the product spacetime $\mathcal{M}_4 = \mathbb{R}^{1,1}\times\Sigma_2$. We use the Schouten tensor $S_{\mu\nu}$ as a notational convenience that compactifies the Weyl decomposition formula.

In four dimensions, the Weyl tensor can be written in terms of the Riemann and Schouten tensors as
\begin{equation}
C_{\mu\nu\rho\sigma}
=
R_{\mu\nu\rho\sigma}
-
2 \left(
g_{\mu[\rho}S_{\sigma]\nu}
-
g_{\nu[\rho}S_{\sigma]\mu}
\right),
\label{eq:Weyl_from_Schouten}
\end{equation}
where antisymmetrization is defined as $A_{[\mu\nu]}=\tfrac{1}{2}(A_{\mu\nu}-A_{\nu\mu})$, and the Schouten tensor has the form
\begin{equation}
S_{\mu\nu}
=
\frac{1}{2}
\left(
R_{\mu\nu}
-
\frac{R}{6}g_{\mu\nu}
\right).
\label{eq:Schouten_def}
\end{equation}
Using the curvature results \eqref{eq:4d_Ricci} and \eqref{eq:4d_scalar_equals_2d} of the previous section, the nonzero components are computed directly. For the transverse block, since $R_{ab}=\frac{R^{(2)}}{2}h_{ab}$ and $g_{ab}=h_{ab}$:
\begin{equation}
S_{ab}
=\frac{1}{2}\!\left(
\frac{R^{(2)}}{2}h_{ab}
-\frac{R^{(2)}}{6}h_{ab}
\right)
=\frac{R^{(2)}}{6}\,h_{ab}.
\label{eq:Sab}
\end{equation}
For the longitudinal $uv$-component, since $R_{uv}=0$ and $g_{uv}=1$:
\begin{equation}
S_{uv}
=\frac{1}{2}\!\left(
0-\frac{R^{(2)}}{6}\cdot 1
\right)
=-\frac{R^{(2)}}{12}.
\label{eq:Suv}
\end{equation}
All remaining components vanish,
\begin{equation}
S_{uu}=S_{vv}=S_{ua}=S_{va}=0,
\label{eq:Szero}
\end{equation}
since both $R_{\mu\nu}$ and $g_{\mu\nu}$ vanish for those index combinations. The Schouten tensor thus inherits the block structure of the Ricci tensor: the transverse block is proportional to $h_{ab}$, the only nonzero longitudinal component is $S_{uv}$, and all mixed components vanish.

\paragraph{Nonvanishing Weyl components.}
Substituting \eqref{eq:Sab}--\eqref{eq:Szero} into
\eqref{eq:Weyl_from_Schouten}, and recalling from \eqref{eq:4d_Riemann_equals_2d}
that $R_{\mu\nu\rho\sigma}$ has support only in the purely transverse sector,
a direct evaluation of \eqref{eq:Weyl_from_Schouten} yields three
algebraically independent nonzero types.
 
\begin{itemize}
\item \emph{Mixed longitudinal-transverse components.}
Setting $\mu=u,\,\nu=a,\,\rho=v,\,\sigma=b$ in \eqref{eq:Weyl_from_Schouten},
with $R_{uavb}=0$, the only surviving Schouten contributions are
$g_{uv}S_{ab}=\frac{R^{(2)}}{6}h_{ab}$ and
$g_{ab}S_{uv}=-\frac{R^{(2)}}{12}h_{ab}$:
\begin{align}
C_{uavb}
&= 0
-\bigl(
\underbrace{g_{uv}}_{1}S_{ab}
-\underbrace{g_{ub}}_{0}S_{av}
-\underbrace{g_{av}}_{0}S_{ub}
+g_{ab}S_{uv}
\bigr)
\notag\\
&= -\frac{R^{(2)}}{6}\,h_{ab}+\frac{R^{(2)}}{12}\,h_{ab}
= -\frac{R^{(2)}}{12}\,h_{ab}.
\label{eq:Cuavb}
\end{align}
 
\item \emph{Purely longitudinal component.}
Setting $\mu=u,\,\nu=v,\,\rho=u,\,\sigma=v$ in \eqref{eq:Weyl_from_Schouten},
with $R_{uvuv}=0$ and $g_{uu}=g_{vv}=0$, only the cross-terms survive:
\begin{align}
C_{uvuv}
&= 0
-\bigl(
\underbrace{g_{uu}}_{0}S_{vv}
-\underbrace{g_{uv}}_{1}S_{vu}
-\underbrace{g_{vu}}_{1}S_{uv}
+\underbrace{g_{vv}}_{0}S_{uu}
\bigr)
\notag\\
&= S_{uv}+S_{uv}
= 2\!\left(-\frac{R^{(2)}}{12}\right)
= -\frac{R^{(2)}}{6}.
\label{eq:Cuvuv}
\end{align}
 
\item \emph{Purely transverse components.}
Setting all indices transverse in \eqref{eq:Weyl_from_Schouten},
with $g_{ab}=h_{ab}$ and $S_{ab}=\frac{R^{(2)}}{6}h_{ab}$:
\begin{align}
C_{abcd}
&= R_{abcd}
-\bigl(
h_{ac}S_{bd}
-h_{ad}S_{bc}
-h_{bc}S_{ad}
+h_{bd}S_{ac}
\bigr)
\notag\\
&= \frac{R^{(2)}}{2}(h_{ac}h_{bd}-h_{ad}h_{bc})
-\frac{R^{(2)}}{6}
\bigl(
2h_{ac}h_{bd}-2h_{ad}h_{bc}
\bigr)
\notag\\
&= \frac{R^{(2)}}{6}(h_{ac}h_{bd}-h_{ad}h_{bc}).
\label{eq:Cabcd}
\end{align}

\end{itemize}
 
All remaining independent components vanish.
The three types are summarised as:
\begin{equation}
C_{uavb} = -\frac{R^{(2)}}{12}\,h_{ab},
\qquad
C_{uvuv} = -\frac{R^{(2)}}{6},
\qquad
C_{abcd} = \frac{R^{(2)}}{6}(h_{ac}h_{bd}-h_{ad}h_{bc}).
\label{eq:Weyl_summary}
\end{equation}
All three types are proportional to the single scalar $R^{(2)}$, confirming that the Weyl tensor is completely encoded in the Gaussian curvature of the transverse factor $\Sigma_2$. This is consistent with the single nonvanishing Newman--Penrose scalar $\Psi_2 = R^{(2)}/12$ and Petrov type~D established in the following subsection.

\subsection{Newman--Penrose Weyl scalars}

The Newman–Penrose scalars are defined as \cite{griffiths-podolsky}
\begin{align}
\Psi_0 &\equiv C_{\mu\nu\rho\sigma}\,\ell^\mu m^\nu \ell^\rho m^\sigma,
\\
\Psi_1 &\equiv C_{\mu\nu\rho\sigma}\,\ell^\mu n^\nu \ell^\rho m^\sigma,
\\
\Psi_2 &\equiv C_{\mu\nu\rho\sigma}\,\ell^\mu m^\nu \bar m^\rho n^\sigma,
\\
\Psi_3 &\equiv C_{\mu\nu\rho\sigma}\,\ell^\mu n^\nu \bar m^\rho n^\sigma,
\\
\Psi_4 &\equiv C_{\mu\nu\rho\sigma}\,n^\mu \bar m^\nu n^\rho \bar m^\sigma .
\end{align}
Because $C_{\mu\nu\rho\sigma}$ is nonzero only for index structure $uavb$, and since $\ell^\mu$ points along $v$ while $n^\mu$ points along $u$, it follows immediately that
\begin{equation}
\Psi_0=\Psi_1=\Psi_3=\Psi_4=0.
\end{equation}
For $\Psi_2$, using \eqref{eq:Cuavb} together with $m^a\bar m^b h_{ab}=1$, we obtain
\begin{equation}
\Psi_2
=
\frac{R^{(2)}}{12}.
\label{eq:Psi2_value}
\end{equation}
Thus the Weyl tensor possesses a single nonvanishing Newman–Penrose scalar.

\subsection{Petrov classification}

\begin{prop}[Petrov type of the product metric]
For the metric \eqref{productmetric}:
\begin{enumerate}
\item If $R^{(2)}\neq 0$, then $\Psi_2\neq 0$ and the spacetime is Petrov type D, with repeated principal null directions
$\ell=\partial_v$ and $n=\partial_u$.
\item If $R^{(2)}=0$, then all $\Psi_i=0$ and the spacetime is conformally flat (Petrov type O).
\end{enumerate}
\end{prop}

\begin{remark}
The algebraic specialty follows directly from the product structure.
Curvature in $\Sigma_2$ generates a Weyl tensor of type~D aligned with the longitudinal null directions. For a round $S^2$ one has $R^{(2)}=2/r_0^2$.
\end{remark}
\section{Einstein--Maxwell--\texorpdfstring{$\Lambda$}{Lambda} system on $\mathbb{R}^{1,1}\times \Sigma_2$}

We now determine under what conditions the product geometry solves the coupled Einstein–Maxwell–$\Lambda$ system
\begin{equation}
G_{\mu\nu}+\Lambda g_{\mu\nu}
=
8\pi G\,T_{\mu\nu},
\label{eq:EinsteinMaxwellLambda}
\end{equation}
together with the Maxwell equations
\begin{equation}
\nabla_\mu F^{\mu\nu}=0,
\qquad
\nabla_{[\mu}F_{\nu\rho]}=0,
\label{eq:Maxwell_eqs}
\end{equation}
and stress tensor
\begin{equation}
T_{\mu\nu}
=
F_{\mu\rho}F_{\nu}{}^{\rho}
-\frac{1}{4}g_{\mu\nu}F_{\rho\sigma}F^{\rho\sigma}.
\label{eq:Tmunu_def}
\end{equation}

As shown in the previous section, the product geometry satisfies $G_{ab}=0$ identically. Consequently the transverse Einstein equations reduce to purely algebraic conditions once matter sources are introduced.
This property is responsible for the simple tuning relations obtained below.

\subsection{Electric ansatz}

Consider the gauge potential
\begin{equation}
A=\alpha v\,\dd u,
\label{eq:A_electric}
\end{equation}
with constant $\alpha$.
The corresponding field strength $F=dA$ is
\begin{equation}
F=\alpha\,\dd v\wedge\dd u,
\qquad
F_{uv}=-\alpha,
\qquad
F_{vu}=+\alpha.
\label{eq:Fuv}
\end{equation}

Since $F_{uv}$ is constant and the metric is independent of $u,v$, the Maxwell equations \eqref{eq:Maxwell_eqs} are automatically satisfied.

\subsubsection*{Invariant and stress tensor}

Raising indices using $g^{uv}=1$ gives
\[
F^{uv}=+\alpha,
\qquad
F^{vu}=-\alpha,
\]
and therefore
\begin{equation}
F^2 \equiv F_{\rho\sigma}F^{\rho\sigma}
=
-2\alpha^2.
\end{equation}

The nonvanishing components of the stress tensor are
\begin{equation}
T_{uv}
=
-\frac{\alpha^2}{2},
\qquad
T_{ab}
=
+\frac{\alpha^2}{2}\,h_{ab}.
\label{eq:T_electric_summary}
\end{equation}

\subsubsection*{Einstein equations and algebraic tuning}

The $(ab)$ components of \eqref{eq:EinsteinMaxwellLambda} give
\begin{equation}
\Lambda h_{ab}
=
8\pi G\left(\frac{\alpha^2}{2}h_{ab}\right),
\end{equation}
which implies
\begin{equation}
\Lambda = 4\pi G\,\alpha^2.
\label{eq:Lambda_alpha}
\end{equation}

The $(uv)$ equation yields
\begin{equation}
-\frac{1}{2}R^{(2)}+\Lambda
=
-4\pi G\,\alpha^2,
\end{equation}
which, together with \eqref{eq:Lambda_alpha}, gives
\begin{equation}
R^{(2)}=16\pi G\,\alpha^2.
\label{eq:R2_alpha}
\end{equation}

Thus the transverse manifold must have constant positive curvature. Both the cosmological constant and the transverse curvature are fixed algebraically by the electric flux.

\begin{prop}[Electric branch]
The product geometry solves the Einstein--Maxwell--$\Lambda$ system with electric flux $\alpha$ if and only if
\[
\Lambda=4\pi G\alpha^2,
\qquad
R^{(2)}=16\pi G\alpha^2.
\]
If $\Lambda=0$, then $\alpha=0$ within this ansatz.
\end{prop}

\subsection{Magnetic and dyonic configurations}

A purely magnetic flux
\begin{equation}
F_{ab}=B\,\eps_{ab},
\qquad
F_{u\mu}=F_{v\mu}=0,
\end{equation}
gives
\begin{equation}
T_{uv}=-\frac{B^2}{2},
\qquad
T_{ab}=+\frac{B^2}{2}h_{ab},
\end{equation}
and therefore the identical tuning relations
\begin{equation}
\Lambda=4\pi G B^2,
\qquad
R^{(2)}=16\pi G B^2.
\end{equation}

For a general dyonic configuration with electric $\alpha$ and magnetic $B$, the stress tensor depends only on
\begin{equation}
Q^2\equiv \alpha^2+B^2.
\end{equation}
The full solution is therefore characterized by the universal algebraic relations
\begin{equation}
\Lambda=4\pi G Q^2,
\qquad
R^{(2)}=16\pi G Q^2.
\label{eq:universal_tuning}
\end{equation}
\section{Unified picture: Nariai, critical flux string, and Bertotti--Robinson}

We now place the critical $\mathbb{R}^{1,1}\times S^2$ solution within the full algebraic family
\[
(\dS_2,\mathbb{R}^{1,1},\AdS_2)\times S^2.
\]
All three branches arise from the same reduced Einstein--Maxwell--$\Lambda$ system once both factors are allowed to carry constant curvature. In this sense, the flat branch is not an isolated special case but one distinguished member of a single unified family.

\subsection{Constant-curvature product ansatz}

Consider the general product metric
\begin{equation}
\dd s^2=\dd s^2(M_2)+r_0^2\,\dd\Omega_2^2,
\label{eq:general_product}
\end{equation}
where $M_2$ is a two-dimensional Lorentzian space of constant curvature $K_L$, $\dd\Omega_2^2$ is the unit round metric on $S^2$, and $r_0$ is the transverse sphere radius. This ansatz extends the flat product geometry of the previous sections to the case in which the longitudinal factor is allowed to have nonzero constant curvature.

For a direct product of constant-curvature manifolds, the Ricci tensor splits as
\begin{equation}
R_{\alpha\beta}=K_L g_{\alpha\beta},
\qquad
R_{ab}=\frac{1}{r_0^2}\,g_{ab},
\label{eq:Ricci_factors}
\end{equation}
where Greek indices refer to $M_2$ and $(a,b)$ to $S^2$.

A homogeneous dyonic Maxwell field compatible with the symmetries has stress tensor
\begin{equation}
T_{\alpha\beta}
=
-\frac{Q^2}{2r_0^4}\,g_{\alpha\beta},
\qquad
T_{ab}
=
+\frac{Q^2}{2r_0^4}\,g_{ab},
\label{eq:T_factors}
\end{equation}
where $Q^2=\alpha^2+B^2$ is the total electric--magnetic charge squared.

\subsection{Algebraic reduction of the Einstein equations}

Substituting \eqref{eq:Ricci_factors} and \eqref{eq:T_factors} into the Einstein--Maxwell--$\Lambda$ equations, one finds that the full field equations reduce to the two algebraic conditions
\begin{align}
K_L &= \Lambda-\frac{4\pi GQ^2}{r_0^4},
\label{eq:KL_eq}\\
\frac{1}{r_0^2} &= \Lambda+\frac{4\pi GQ^2}{r_0^4}.
\label{eq:r0_eq}
\end{align}
The first relation makes the geometric meaning of $K_L$ completely transparent: the Lorentzian curvature is determined by the competition between the cosmological constant and the Maxwell flux. As the relative strength of these two contributions varies, $K_L$ continuously passes through positive, zero, and negative values.

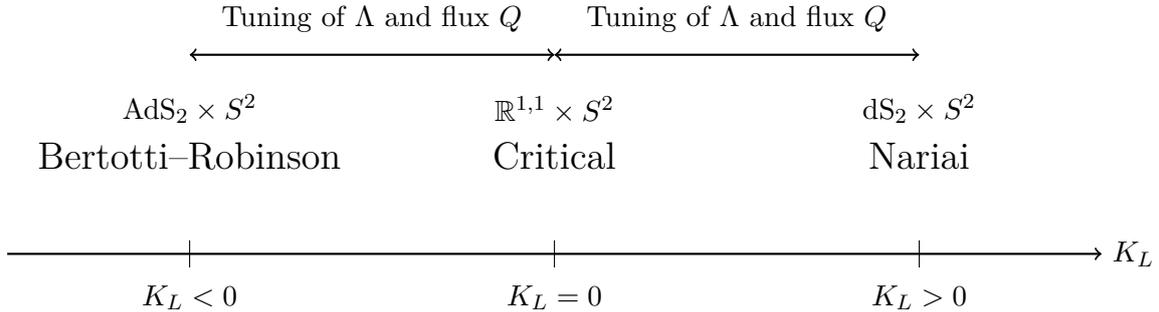
\begin{figure}[t]
\centering
\begin{tikzpicture}[scale=1.2]

\draw[->, thick] (-6,0) -- (6,0) node[right] {$K_L$};

\draw (-4,0.15) -- (-4,-0.15);
\draw (0,0.15) -- (0,-0.15);
\draw (4,0.15) -- (4,-0.15);

\node[below=8pt] at (-4,0) {$K_L<0$};
\node[below=8pt] at (0,0) {$K_L=0$};
\node[below=8pt] at (4,0) {$K_L>0$};

\node[above=1.0cm] at (-4,0) {\large Bertotti--Robinson};
\node[above=1.6cm] at (-4,0) {$\AdS_2 \times S^2$};

\node[above=1.0cm] at (0,0) {\large Critical};
\node[above=1.6cm] at (0,0) {$\mathbb{R}^{1,1} \times S^2$};

\node[above=1.0cm] at (4,0) {\large Nariai};
\node[above=1.6cm] at (4,0) {$\dS_2 \times S^2$};

\draw[<->, thick] (-4,2.2) -- (0,2.2);
\draw[<->, thick] (0,2.2) -- (4,2.2);

\node at (-2,2.6) {Tuning of $\Lambda$ and flux $Q$};
\node at (2,2.6) {Tuning of $\Lambda$ and flux $Q$};

\end{tikzpicture}
\caption{Algebraic interpolation between the Nariai, critical, and Bertotti--Robinson branches. The effective Lorentzian curvature $K_L$ changes sign as the balance between $\Lambda$ and the conserved flux $Q$ varies. The flat branch $K_L=0$ is the midpoint at which the longitudinal curvature vanishes. The interpolation is algebraic: it reflects the structure of the reduced field equations rather than a dynamical evolution between solutions.}
\label{fig:interpolation}
\end{figure}

\subsection{The three branches}

The sign of $K_L$ distinguishes three geometrically distinct solutions:

\begin{itemize}

\item \textbf{Nariai branch ($K_L>0$):}
$\dS_2\times S^2$ with isometry group $\SO(1,2)\times\SO(3)$.
This branch includes the standard vacuum Nariai solution as a special case. In particular, for vanishing flux $Q=0$ one finds
\[
K_L=\Lambda>0,
\qquad
r_0^{-2}=\Lambda,
\]
which corresponds to the usual $\Lambda$-supported Nariai spacetime.

\item \textbf{Bertotti--Robinson branch ($K_L<0$):}
$\AdS_2\times S^2$ with isometry group $\SO(2,1)\times\SO(3)$. This branch requires a nonvanishing electromagnetic field.
In particular, for vanishing cosmological constant $\Lambda=0$ one obtains
\[
K_L=-r_0^{-2},
\qquad
r_0^2=4\pi GQ^2,
\]
which reproduces the standard Bertotti--Robinson solution.

\item \textbf{Critical branch ($K_L=0$):}
the Maxwell flux string $\mathbb{R}^{1,1}\times S^2$ with isometry group $\ISO(1,1)\times\SO(3)$. This configuration obeys
\begin{equation}
\Lambda=\frac{4\pi GQ^2}{r_0^4},
\qquad
r_0^{-2}=8\pi GQ^2,
\label{eq:critical_conditions}
\end{equation}
where $Q\ne0$. Equivalently,
\begin{equation}
\Lambda=\frac{1}{2r_0^2}.
\label{eq:critical_lambda_r0}
\end{equation}
This configuration represents the balanced case in which the cosmological constant and the Maxwell flux exactly cancel in the longitudinal curvature, yielding a flat $\mathbb{R}^{1,1}$ factor.

\end{itemize}
\medskip

\noindent
Among these three branches, the critical solution is distinguished not merely as a limiting case but as the unique point at which the Lorentzian curvature $K_L$ changes sign. At $K_L=0$, the longitudinal sector becomes exactly flat and the reduced Einstein equations undergo a qualitative degeneracy: the constraint fixing the longitudinal curvature disappears, allowing Brinkmann-type null deformations. This degeneracy can be traced directly to the reduced field equations, where the algebraic constraint on $K_L$ ceases to impose an independent restriction at the critical point. This identifies the $K_L=0$ branch as the structural midpoint of the family.

The Nariai and Bertotti--Robinson geometries therefore arise as distinguished limits of the same reduced Einstein--Maxwell--$\Lambda$ system, rather than as fundamentally separate solutions. The critical branch provides the geometric and algebraic interpolation between them. This structure is reflected at the level of isometries,
\[
\SO(1,2)\times\SO(3) \;\longrightarrow\; \ISO(1,1)\times\SO(3) 
\;\longrightarrow\; \SO(2,1)\times\SO(3),
\]
as the relative balance between the cosmological constant $\Lambda$ and the Maxwell flux $Q$ is varied. The transverse $\SO(3)$ symmetry of the two-sphere is preserved throughout, while the longitudinal isometry group transitions between the constant-curvature geometries $\dS_2$, $\mathbb{R}^{1,1}$, and $\AdS_2$. The intermediate Poincaré symmetry $\ISO(1,1)$ signals the degeneration of curvature into a flat geometry, in direct correspondence with the loss of the longitudinal curvature constraint 
at $K_L=0$. This symmetry transition underlies the appearance of a degenerate Kundt structure in the critical branch, where the longitudinal sector admits nontrivial null deformations propagating along aligned null directions.

\section{Brinkmann $pp$-wave deformation and rigidity on a round sphere}

We now examine whether the critical background admits null deformations that preserve the product structure of the transverse sphere while allowing nontrivial longitudinal dynamics. The null vector field defining the deformation is covariantly constant; therefore, the deformed spacetime is a pp-wave. 

\subsection{Metric deformation}

Consider the Brinkmann-type deformation of the critical metric:
\begin{equation}
\dd s^2
=
2\,\dd u\,\dd v
+
2V(u,x)\,\dd u^2
+
r_0^2\,\dd\Omega_2^2,
\label{eq:pp_wave_metric}
\end{equation}
where $V$ is independent of $v$, and the same homogeneous dyonic Maxwell flux is retained.

This metric belongs to the Kundt class. In particular, $\ell=\partial_v$ remains null, geodesic, shear-free, twist-free, and expansion-free; indeed, in the Brinkmann form it remains covariantly constant. The deformation introduces a single additional function $V(u,x)$ along the preferred null direction.

\subsubsection*{Einstein equations}

The only modified curvature component is $R_{uu}$. A direct computation gives
\begin{equation}
R_{uu}
=
-\Delta_{\Sigma_2} V,
\end{equation}
where $\Delta_{\Sigma_2}$ is the Laplacian on the transverse space.

Since the homogeneous Maxwell stress tensor satisfies $T_{uu}=0$, the $uu$ component of the Einstein equations reduces to
\begin{equation}
R_{uu}=0.
\end{equation}
Hence the deformation function obeys the harmonic equation
\begin{equation}
\Delta_{\Sigma_2} V = 0.
\label{eq:Laplace_V}
\end{equation}

Crucially, this equation contains no algebraic curvature term. It is therefore a genuinely differential constraint. This degeneracy occurs precisely because the longitudinal curvature $K_L$ vanishes in the critical branch.

\subsection{Rigidity on a round sphere}

For the round sphere $S^2$, the scalar Laplacian has eigenvalues
\[
-\Delta_{S^2} Y_{\ell m}
=
\frac{\ell(\ell+1)}{r_0^2} Y_{\ell m},
\qquad \ell=0,1,2,\dots.
\]
The only smooth harmonic scalars are the $\ell=0$ modes, which are constant over the sphere. Hence
\begin{equation}
V(u,x)=V(u).
\label{eq:V_of_u}
\end{equation}
However, a function of $u$ alone can be removed by the coordinate redefinition
\begin{equation}
v
\;\longmapsto\;
v-\int^{u}V(u')\,\dd u'.
\label{eq:remove_V}
\end{equation}
Therefore:

\begin{prop}[Rigidity on $S^2$]
For the round sphere, every smooth Brinkmann-type deformation of the critical background is pure gauge. No nontrivial $pp$-wave excitations exist on $\mathbb{R}^{1,1}\times S^2$.
\end{prop}

One can show that the above result is intact for the $\dS_2\times S^2$ and $\AdS_2\times S^2$ cases.

\section{Physical interpretation and discussion}

The critical solution is neither a black-hole spacetime nor asymptotically flat. Instead, it describes a homogeneous electromagnetic flux geometry whose intrinsic length scale is determined entirely by the conserved flux threading the transverse sphere. The geometry is everywhere regular, compact in the transverse directions, and noncompact and flat along the longitudinal $\mathbb{R}^{1,1}$ factor. In this sense, the spacetime should be interpreted as a four-dimensional \emph{flux string} or \emph{fluxbrane}: a configuration in which gauge-field energy density supports curvature in directions orthogonal to a homogeneous null worldvolume. The Maxwell field does not source a localized object; rather, it permeates the spacetime and stabilizes the transverse sphere through an algebraic balance with the cosmological constant.

\subsection*{Relation to extremal geometries}

Although no horizon is present, the solution shares structural features with extremal configurations:

\begin{itemize}
\item an algebraic relation between charge and transverse radius,
\item a marginal balance between curvature and flux,
\item constant scalar curvature invariants,
\item alignment of principal null directions,
\item enhanced symmetry relative to generic Einstein--Maxwell solutions.
\end{itemize}

The charge–radius relation resembles the attractor behavior familiar from extremal black holes: the transverse size is not freely adjustable but is fixed by the conserved flux. However, unlike the attractor mechanism in $\AdS_2\times S^2$ near-horizon geometries, here the Lorentzian factor is exactly flat.
The $\SO(2,1)$ symmetry of $\AdS_2$ is replaced by the Poincaré group $\ISO(1,1)$, and no near-horizon interpretation applies.

\subsection*{Symmetry interpolation}

The three members of the family exhibit a continuous symmetry pattern:
\[
\SO(1,2)\times\SO(3)
\;\longrightarrow\;
\ISO(1,1)\times\SO(3)
\;\longrightarrow\;
\SO(2,1)\times\SO(3),
\]
corresponding to Nariai, critical, and Bertotti--Robinson geometries. The critical solution sits precisely at the transition where the Lorentzian curvature changes sign. It is therefore a geometric and algebraic midpoint in parameter space.

\subsection*{Algebraic structure and almost universality}

Another remarkable feature is the algebraic simplicity of the curvature. The Riemann tensor splits cleanly into longitudinal and transverse parts, and all scalar curvature invariants are constant. The Weyl tensor is of Petrov type~D in the undeformed background, with repeated principal null directions aligned along the longitudinal factor.

Because the curvature has only a small number of independent scalars, any symmetric rank-two tensor constructed polynomially from the Riemann tensor reduces to a linear combination of the metric and the
Maxwell stress tensor. Consequently, a broad class of higher-curvature metric theories admit the same background once the same algebraic tuning conditions are imposed. In this precise sense, the critical flux string is an \emph{almost universal} spacetime.

The critical $\mathbb{R}^{1,1}\times S^2$ configuration is best understood not as a limiting black-hole geometry, but as a homogeneous flux geometry interpolating algebraically between the Nariai and Bertotti--Robinson spacetimes. Its defining features — flat longitudinal geometry, midpoint degeneracy, and algebraic rigidity — make it a natural four-dimensional analog of fluxbrane solutions familiar in higher-dimensional gravity.

It is precisely this combination of simplicity, flexibility, and structural rigidity that singles out the critical branch within the $(\dS_2,\mathbb{R}^{1,1},\AdS_2)\times S^2$ family.

\section{Conclusions and further directions}

In this work we have presented an analysis of a homogeneous Einstein--Maxwell--$\Lambda$ configuration of the form
\[
\mathbb{R}^{1,1}\times \Sigma_2,
\]
with particular emphasis on the critical branch $\mathbb{R}^{1,1}\times S^2$ that interpolates algebraically between the Nariai and Bertotti--Robinson geometries.

The essential results may be summarized as follows.

\medskip

\noindent
\textbf{(1) Algebraic unification.}
The Nariai, critical, and Bertotti--Robinson geometries form a single algebraic family distinguished solely by the sign of the Lorentzian curvature $K_L$. The critical branch $K_L=0$ is the precise midpoint at which the longitudinal curvature changes sign. The transverse sphere radius is fixed algebraically by the conserved
Maxwell flux and cosmological constant, mirroring the charge--radius relation familiar from extremal
near-horizon geometries.

\medskip

\noindent
\textbf{(2) Algebraic structure and Petrov type.}
The undeformed background is Petrov type~D with repeated principal null directions aligned along the longitudinal factor. 

\medskip

\noindent
\textbf{(3) Almost universality.}
One of the most striking features of the critical flux string is its algebraic simplicity.

The spacetime possesses:
\begin{itemize}
\item constant scalar curvature invariants (CSI),
\item aligned traceless Ricci tensor,
\item a degenerate Kundt null direction,
\item a Riemann tensor expressible entirely in terms of
      $g_{ab}$ and the Maxwell stress tensor.
\end{itemize}

Because of this structure, any symmetric rank-two tensor constructed polynomially from the Riemann tensor
(without derivatives) reduces on this background to
\[
\mathcal{T}_{\mu\nu}
=
\alpha\, g_{\mu\nu}
+
\beta\, T^{(\text{Maxwell})}_{\mu\nu},
\]
with constant coefficients $\alpha$ and $\beta$.

Consequently, the field equations of a broad class of metric-based higher-curvature gravity theories ---
including $f(\mathrm{Riemann})$ models and general quadratic curvature actions --- collapse on this background to the same algebraic tuning relations between $\Lambda$, transverse curvature, and conserved flux. In this precise sense, the critical flux string (and its aligned Kundt deformations) constitutes an \emph{almost universal} solution \cite{GST-AUM2026}. 

\bigskip

In summary, the critical $\mathbb{R}^{1,1}\times S^2$ flux geometry is not merely an algebraic curiosity. It occupies a distinguished position within the product family linking the Nariai and Bertotti--Robinson spacetimes.
Its Kundt structure and algebraic simplicity make it a natural candidate for universality across a broad class of metric-based gravitational theories.


\end{document}